\documentclass[12pt]{article}

\usepackage{scicite}

\usepackage{times}

\usepackage{graphicx}
\usepackage{hyperref}

\newenvironment{sciabstract}{%
\begin{quote} \bf}
{\end{quote}}

\newcommand{\MSun}{M_\odot}

\newcommand{\onesigrange}[3]{\ensuremath{#1^{+#2}_{-#3}}}
\newcommand{\alpharangeone}{\onesigrange{2.05}{0.14}{0.13}}
\newcommand{\alpharangetwo}{\onesigrange{2.05}{0.14}{0.13}}
\newcommand{\alpharangethree}{\onesigrange{2.11}{0.19}{0.17}}
\newcommand{\alpharangefour}{\onesigrange{2.15}{0.13}{0.13}}

\title{Comment on ``An excess of massive stars in the local 30 Doradus starburst''}

\author{Will M. Farr,$^{1,2\ast}$ Ilya Mandel$^{1,3\ast}$\\
\normalsize{$^1$Institute of Gravitational Wave Astronomy and School of Physics and Astronomy,}\\
\normalsize{University of Birmingham, Birmingham, B15 2TT, United Kingdom}\\
\normalsize{$^2$Center for Computational Astrophysics, Flatiron Institute}\\
\normalsize{162 Fifth Avenue, New York NY 10010, United States} \\
\normalsize{$^3$Monash Centre for Astrophysics, School of Physics and Astronomy,}\\
\normalsize{Monash University, Clayton, Victoria 3800, Australia}\\
\normalsize{$^\ast$E-mail: wmfarr@star.sr.bham.ac.uk, imandel@star.sr.bham.ac.uk}
}

\date{}

\begin{document}


\baselineskip24pt


\maketitle

\begin{sciabstract}
Schneider et al.~(Reports, 5 January 2018, p.~69) used an ad hoc statistical method in their calculation of the stellar initial mass function. Adopting an improved approach, we reanalyse their data and determine a power law exponent of $\alpharangeone$. Alternative assumptions regarding data set completeness and the star formation history model can shift the inferred exponent to $\alpharangethree$ and $\alpharangefour$, respectively.
\end{sciabstract}

Schneider et al.\cite{Schneider:2018} use spectroscopic observations of young massive stars in
the 30 Doradus region of the Large Magellanic Cloud to infer a
shallower-than-expected stellar initial mass function (IMF) with a power-law exponent of $\alpha=1.90^{+0.37}_{-0.26}$, in contrast to the Salpeter exponent of $2.35$\cite{Salpeter:1955}.  They estimate the ages and masses of individual
stars with the BONNSAI Bayesian code \cite{Schneider:2017}, then obtain
an overall mass distribution by effectively adding together the posterior
probability density functions of individual stars.  There is no statistical
meaning to a distribution obtained in this way, which does not represent the
posterior probability density function on the mass distribution.

Hierarchical Bayesian inference provides the statistically justified solution to
this problem \cite{Hogg:2010}.  Mandel\cite{Mandel:2010stat} specifically considered
inference on a mass distribution given a sample of uncertain measurements, and
we use a similar methodology here.  We interpret the Schneider et al.\cite{Schneider:2018}
inference on individual masses and ages as independent Gaussian likelihoods for
the logarithm of the mass and the age, with parameters fixed by matching the mean
parameter to the peak and the standard deviation parameter to the 68\% width of
the individual stellar distributions in the Schneider et al.\cite{Schneider:2018} data.

For our fiducial analysis we model the star formation history as a truncated Gaussian distribution, and
generally find a star formation history similar to that in Schneider et al.\cite{Schneider:2018}, with
the star formation rate in 30 Doradus peaking about 4 million years ago.  We impose broad priors
on the power--law exponent and the mean and standard deviation of the star
formation Gaussian.    We use the Hamiltonian Monte Carlo sampler Stan
\cite{STAN} to efficiently address the high-dimensional hierarchical problem
with free parameters for each star's actual mass and age in addition to
the IMF exponent and the mean and standard deviation of the star formation
history.

\begin{figure}
\includegraphics[width=\columnwidth]{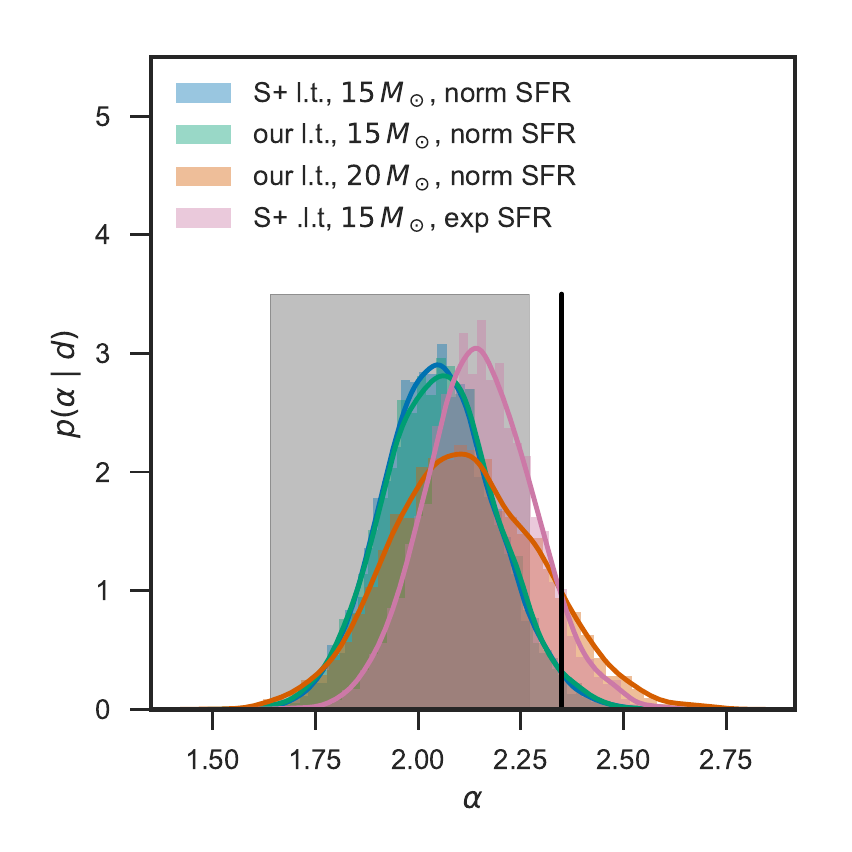}
    		\caption{{\bf The posterior on the IMF power-law exponent $\alpha$ is inferred from the observations $d$.}  See text for details on the four models (top to bottom in the legend): Blue: Schneider et al.\cite{Schneider:2018} [S+] stellar lifetimes [l.t.], survey completeness for $M \geq 15 \, \MSun$, and Gaussian star formation history model; green: same, but with our lifetime fit; orange: as green, but with completeness for $M \geq 20 \, \MSun$; pink: as blue, but with a double-exponential star formation history model.  The Salpeter power-law exponent is $-\alpha=-2.35$ \cite{Salpeter:1955}, indicated by a vertical black line.  The 68.3\% range of power-law exponents derived by Schneider et al.\cite{Schneider:2018} is shaded in grey.}\label{fig:IMF}
\end{figure}

Figure \ref{fig:IMF} shows the inferred power-law exponent of the IMF.   We use the Schneider et al.\cite{Schneider:2018} fit to stellar lifetimes and assume
that their data set is complete above 15 solar masses; that is, we select only
those stars whose observed mass is above $15 \, M_\odot$
\cite{Loredo:2004,BBH:O1}.  We find
an exponent of $\alpha=\alpharangeone$ where the quoted value corresponds to the median
of the posterior distribution and the range to the 16th and 84th percentiles
(i.e.\ the symmetric 68\% credible interval).  Hierarchical Bayesian modelling steepens the preferred IMF
slope; our median $\alpha$ value lies about $1\, \sigma$ above the preferred value from
Schneider et al.\cite{Schneider:2018}.  More significantly, this analysis narrows the uncertainty interval by more than a factor of 2.

The analysis above uses the same assumptions as Schneider et al.\cite{Schneider:2018}.  Below, we consider the impact of three additional  assumptions: the stellar lifetime fit, the choice of the completeness limit, and the model for the star formation history.

We performed an independent fit to the main sequence lifetimes
$\tau_{MS}$ of non-rotating massive stars of mass $M$ as modelled by Brott et al.\cite{Brott:2011} and K\"{o}hler et al.\cite{Kohler:2015}:
\begin{equation}
\ln \frac{\tau_{MS} (M)}{\textrm{Myr}} = 9.1973 - 3.8955 \ln\frac{M}{M_\odot} 
+ 0.6107 \left(\ln\frac{M}{M_\odot} \right)^2 - 0.0332 \left(\ln\frac{M}{M_\odot}\right)^3.
\end{equation}
Following Schneider et al.\cite{Schneider:2018}, we increased the ``observable''
lifetime of a star by 10\% beyond its main-sequence lifetime to account for
helium burning.  We find that this alternative fit does not affect the inferred IMF, yielding the same power-law exponent $\alpha=\alpharangetwo$.

The inferred power-law exponent is somewhat sensitive to
the choice of the cutoff mass for survey completeness.  The data of Schneider et al.\cite{Schneider:2018} show a relative scarcity of stars between 15 $M_\odot$ and 20 $M_\odot$; changing the mass cutoff from $15 M_\odot$ to $20 M_\odot$ further steepens the inferred exponent to $\alpha=\alpharangethree$.    However, these fluctuations are within the expected statistical variation based on the sample size, as confirmed with posterior predictive
checking.  In particular, there is no statistical evidence against the claim of Schneider et al.\cite{Schneider:2018} that
the survey is complete for $M \geq 15 \, \MSun$.

Finally, we considered an alternative star formation history model -- a double exponential with three free parameters: the time of the peak of the star formation rate, and the (possibly different) decay constants before and after the peak.  This model allows for a sharper peak and longer tails than a Gaussian.  This star formation rate history model is consistent with the data, as tested with posterior predictive checking (see below).  However, it yields a power-law exponent $\alpha=\alpharangefour$, almost  $1\, \sigma$ steeper than for our fiducial analysis.  This indicates that the inferred IMF is sensitive to the systematics of the assumed star formation history model.

We also considered the possibility that the IMF power law has an additional break at higher masses, allowing for three free parameters: the mass at which the break happens and the exponent below and above the break.  However, we find that the data do not constrain the parameters of this more general model, and there is no statistical preference for a broken power-law model.

\begin{figure}
    		    A.\\		\includegraphics[width=0.5\columnwidth]{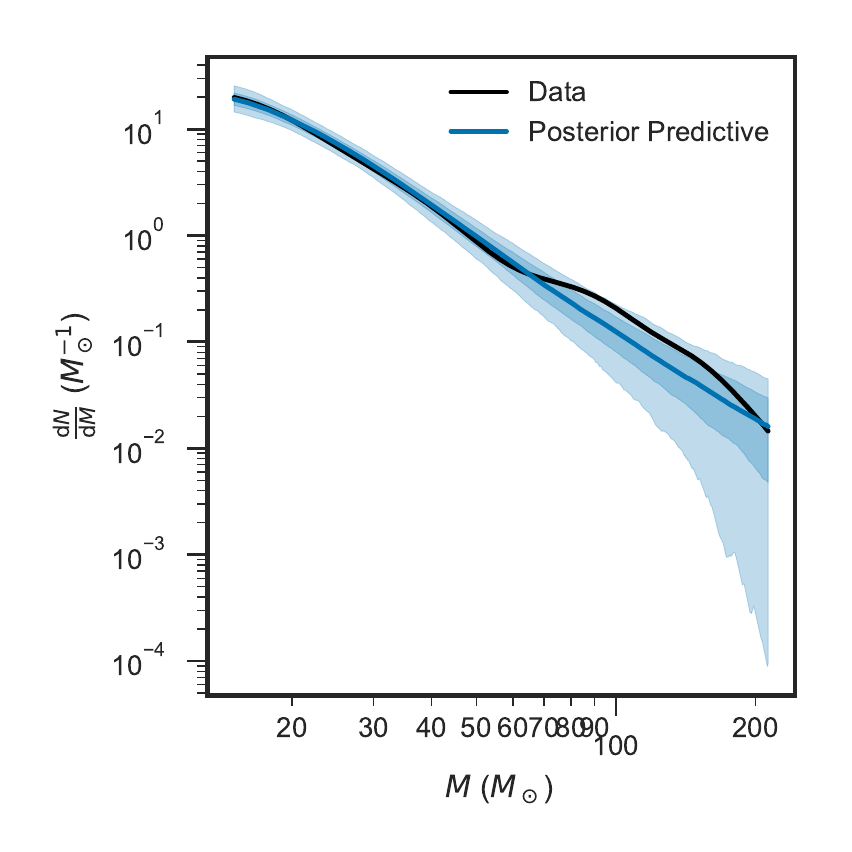}\\
               B.\\  \includegraphics[width=0.5\columnwidth]{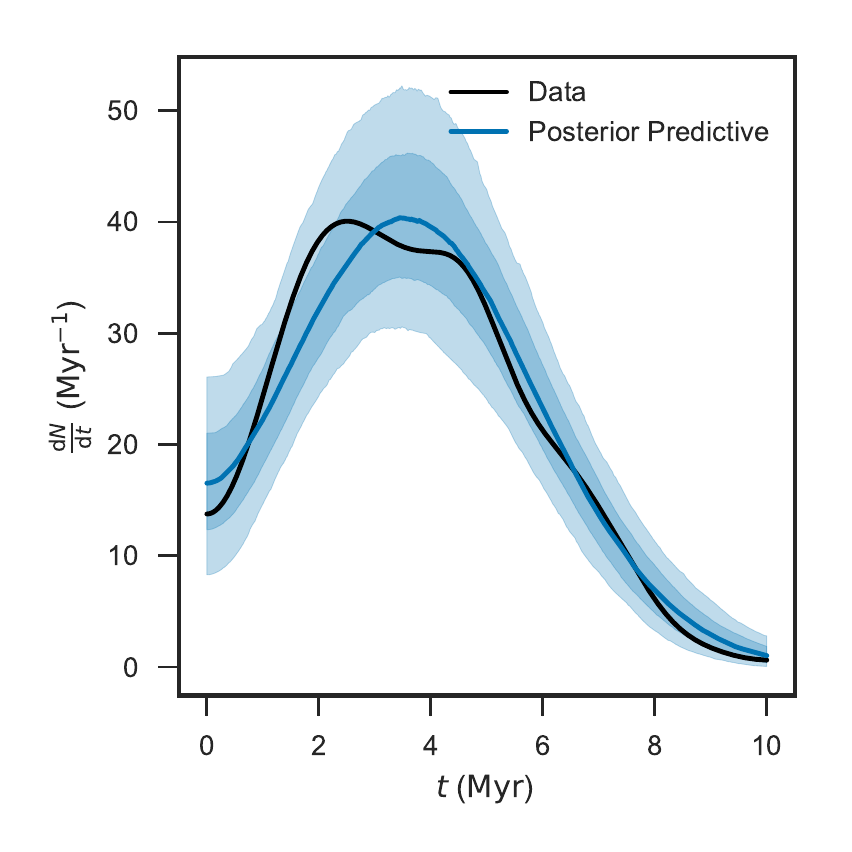}
    		\caption{{\bf Posterior predictive checking demonstrates that the observed data are consistent with being drawn from our model.} The observed distribution of (maximum likelihood) masses $M$ (A) and ages $t$ (B) are shown as black curves; distributions of mass and age from synthetic data drawn from our fitted model (i.e.\ the posterior predictive distribution) are shown as blue curves (median) and shaded blue regions (68\%, and 95\% credible intervals), respectively.}\label{fig:PPC}
\end{figure}

We confirmed the stability of our conclusions with posterior predictive checking.
Figure \ref{fig:PPC} shows the distribution of observed masses and ages (i.e.,
the peak of the likelihood) from the Schneider et al.\cite{Schneider:2018} data overlain on the
range of mass and age distributions that would be observed from a large number
of data sets drawn according to our fitted fiducial IMF model.  The data are consistent with being drawn from our model.  We have also confirmed that all of our models yield predictions for the numbers of stars heavier than $30 M_\odot$ and $60 M_\odot$ that are consistent with observations.

We find that we can substantially reduce the statistical uncertainty in the IMF by applying an improved statistical analysis to the observations of young massive stars in 30 Doradus.  However, the systematics from modelling uncertainties, such as the assumed star formation history model, can potentially shift the inferred power-law exponent by more than the statistical uncertainty.   Furthermore, we adopted the mass and age posteriors for individual stars directly from Schneider et al.\cite{Schneider:2018}.  Imperfect stellar models or the inclusion of other complicating factors described by Schneider et al.\cite{Schneider:2018} (rotation, mass transfer, mergers, etc.) introduce further systematic uncertainty that could again shift the inferred IMF exponent.  The combination of these factors makes it very challenging to infer the precise shape of the IMF even when a data set as good as that obtained by Schneider et al.\cite{Schneider:2018} is available.

{\bf Acknowledgments:} We thank Schneider et al.\cite{Schneider:2018} for making available for further study and analysis the data on which their conclusions are based , and F.~Schneider personally for very useful discussions.  This analysis made use of \texttt{PyStan} \cite{STAN}, \texttt{astropy} \cite{astropy}, \texttt{numpy} \cite{numpy}, \texttt{scipy} \cite{scipy}, \texttt{matplotlib} \cite{matplotlib}, and \texttt{seaborn} \cite{seaborn} Python libraries.

{\bf Funding:} WMF and IM are partially supported by STFC.

{\bf Author contributions:} WMF and IM are jointly responsible for all aspects of this work.

{\bf Competing interests:} None.

{\bf Data and materials availability:} The code and \LaTeX source used to prepare this document are publicly available under an open-source MIT license at \url{https://github.com/farr/30DorIMF}.

\bibliographystyle{Science}
\bibliography{Mandel}
\end{document}